\documentclass{article}
\usepackage{spconf,amsmath,graphicx}
\usepackage{cite}
\usepackage{amsmath,amssymb,amsfonts}
\usepackage{textcomp}
\usepackage[table,xcdraw]{xcolor}
\usepackage{multirow}
\usepackage{acronym} 
\usepackage{svg}
\usepackage[hypcap=true]{subcaption, caption}
\usepackage{graphicx}
\usepackage{adjustbox}

\usepackage[citecolor=black,
            colorlinks=true,
            linkcolor=black,
            urlcolor  = black,
            pdfpagelabels={true},
            pdftitle={Deformable Temporal Convolutional Networks for Monaural Noisy Reverberant Speech Separation},
            pdfauthor={William Ravenscroft, Stefan Goetze, Thomas Hain},
            pdfkeywords={speech separation, deformable convolution, dynamic neural networks},
            pdfsubject={speech separation, deformable convolution, dynamic neural networks}]
            {hyperref}
\usepackage{orcidlink}
\acrodef{AC}    {acoustic conditions}
\acrodef{ASR}   {automatic speech recognition}
\acrodef{BLSTM} {bidirectional long short term memory}
\acrodef{CB}    {convolutional block}
\acrodefplural{CB}[CBs]{convolutional blocks}
\acrodef{Conv-TasNet} {convolutional \ac{TasNet}}
\acrodef{CSM}   {clean speech mixtures}
\acrodef{cLN}   {channelwise layer normalization}
\acrodef{DAE}   {denoising autoencoder}
\acrodef{D-Conv}{depthwise convolution}
\acrodef{DD-Conv}{deformable depthwise convolution}
\acrodef{DDS-Conv}{deformable depthwise-separable convolution}
\acrodef{DS-Conv}{depthwise-separable convolution}
\acrodef{DM}    {dynamic mixing}
\acrodef{DL}    {deep learning}
\acrodef{DNN}   {deep neural network}
\acrodef{DPRNN} {dual path recurrent neural network model}
\acrodef{DPTNet}{dual path Transformer network}
\acrodef{DTCN}  {deformable temporal convolutional network}
\acrodef{E2E}   {end-to-end}
\acrodef{ER}    {early reflection}
\acrodefplural{ER}[ERs]{early reflections}
\acrodef{gLN}   {global layer normalization}
\acrodef{FLOP/s} {floating point operations per second}
\acrodef{LSTM}  {long short term memory}
\acrodef{LR}    {late reflection}
\acrodefplural{LR}[LRs]{late reflections}
\acrodef{MACs}  {mutiply-acccumulate operations}
\acrodef{MHA}   {Multihead Attention}
\acrodef{MOS}   {Mean Opinion Score}
\acrodef{MR}    {mask refinement}
\acrodef{NMF}   {non-negative matrix factorization}
\acrodef{NSM}   {noisy speech mixture}
\acrodef{NRSM}  {noisy reverberant speech mixture}
\acrodef{PESQ}  {perceptual evaluation of speech quality}
\acrodef{PIT}   {permutation invariant training}
\acrodef{PM}    {post-masking}
\acrodef{P-Conv}{pointwise convolution}
\acrodef{PReLU} {parametric \ac{ReLU}}
\acrodef{ReLU}  {rectified linear unit}
\acrodef{RF}    {receptive field}
\acrodef{RIR}   {room impulse response}
\acrodef{RIRs}   {room impulse responses}
\acrodefplural{RIR}[RIRs]{room impulse responses}
\acrodef{RSM}   {reverberant speech mixture}
\acrodef{SA}    {speed augmentation}
\acrodef{SC}    {skip connection}
\acrodef{SISDR} {scale-invariant signal-to-distortion ratio}
\acrodef{SDR}   {signal-to-distortion ratio}
\acrodef{SNR}   {signal-to-noise ratio}
\acrodef{SRMR}  {speech-to-reverberation modulation energy ratio}
\acrodef{SP}    {signal processing}
\acrodef{STFT}  {short-time Fourier transform}
\acrodef{STOI}  {short-time objective intelligibility}
\acrodef{SOTA}  {state-of-the-art}
\acrodef{SW}    {shared weights}
\acrodef{ESTOI} {extended short-time objective intelligibility}
\acrodef{TasNet}{Time-domain audio separation network}
\acrodef{TCN}   {temporal convolutional network}
\acrodef{UPGMA} {unweighted pair group method with arithmetic mean}
\acrodef{WER}   {word error rate}
\acrodef{WPE}   {weighted prediction error}

\newcommand{\mat}[1]{\mathbf{#1}}
\newcommand{\vek}[1]{\ensuremath{\mathbf{#1}}}    
\newcommand{\vekt}[1]{\ensuremath{\boldsymbol{\mathrm{#1}}}}

\newcommand{\PosInt}{\mathbb{Z}^+}
\newcommand{\Real}{\mathbb{R}}

\def\BibTeX{{\rm B\kern-.05em{\sc i\kern-.025em b}\kern-.08em
    T\kern-.1667em\lower.7ex\hbox{E}\kern-.125emX}}
    
\ninept

\begin{document}
\title{Deformable Temporal Convolutional Networks for Monaural Noisy Reverberant Speech Separation
}

\name{{William Ravenscroft$^{\orcidlink{0000-0002-0780-3303}}$, Stefan Goetze$^{\orcidlink{0000-0003-1044-7343}}$, and Thomas Hain$^{\orcidlink{0000-0003-0939-3464}}$}
\thanks{{This work was supported by the Centre for Doctoral Training in Speech and Language Technologies (SLT) and their Applications funded by UK Research and Innovation [grant number EP/S023062/1].} {This work was also funded in part by 3M Health Information Systems, Inc.}}}

\address{\textit{Department of Computer Science}, 
\textit{{The} University of Sheffield}, Sheffield, United Kingdom \\
\{jwravenscroft1, s.goetze, t.hain\}@sheffield.ac.uk\vspace*{-0.2cm}}



\maketitle

\begin{abstract}
Speech separation models are used for isolating individual speakers in many speech processing applications. Deep learning models have been shown to lead to \ac{SOTA} results on a number of speech separation benchmarks. One such class of models known as \acp{TCN} has shown promising results for speech separation tasks. A limitation of these models is that they have a fixed \ac{RF}. Recent research in speech dereverberation has shown that the optimal \ac{RF} of a \ac{TCN} varies with the reverberation characteristics of the speech signal. In this work deformable convolution is proposed as a solution to allow \ac{TCN} models to have dynamic \acp{RF} that can adapt to various reverberation times for reverberant speech separation. The proposed models are capable of achieving an 11.1~dB average \ac{SISDR} improvement over the input signal on the WHAMR benchmark. A relatively small deformable \ac{TCN} model of 1.3M parameters is proposed which gives comparable separation performance to larger and more computationally complex models.
\end{abstract}


\section{Introduction}
The separation of overlapping speech signals is an area that has been widely studied and which has many applications \cite{Benesty_Source_Separation_2000,Moritz13,FFASRHaebUmbach,ShiHain21}.
Deep learning models have demonstrated impressive results on separating clean speech mixtures \cite{convtasnet,sepformer}. However, this performance still degrades heavily under noisy reverberant conditions \cite{WHAMR}. This performance loss can be alleviated somewhat with careful hyper-parameter optimization but a significant performance gap still exists \cite{IWAENCbestpaper}.

The Conv-TasNet speech separation model has been widely studied and adapted for a number of speech enhancement tasks \cite{convtasnet, beamtasnet, rfield, speakerbeam2}. 
Conv-TasNet generally performs very well on clean speech mixtures with a very low computational cost compared to the 
most performant speech separation models \cite{sepformer,sudormrf,tinysep} on the WSJ0-2Mix benchmark\cite{Isik}.
As such, it is still used in many related areas of research \cite{beamtasnet,speakerbeam2}. 
Recent research efforts in speech separation have focused on producing more resource-efficient models even if they do not produce the most \ac{SOTA} results on separation benchmarks \cite{sudormrf,tinysep}. 
Previous work has investigated adaptations to Conv-TasNet with additional modifications such as multi-scale convolution and gating mechanisms applied to the outputs of convolutional layers but these significantly increase the computational complexity \cite{furcanext}. 
The Conv-TasNet model uses a sequence model known as a \ac{TCN}. 
It was recently shown that the optimal \ac{RF} of \acp{TCN} in dereverberation models varies with reverberation time when the model size is sufficiently large \cite{rfield}. 
Furthermore, it was shown that multi-dilation \ac{TCN} models can be trained implicitly to weight differently dilated convolutional kernels to optimally focus within the \ac{RF} on more or less temporal context according to the reverberation time in the data for dereverberation tasks \cite{wdtcn}, i.e.~for larger reverberation times more weight was given to kernels with larger dilation factors.

In this work deformable depthwise convolutional layers \cite{deformconv,1ddconv,dsconv} are proposed as a replacement for standard depthwise convolutional layers \cite{convtasnet} in \ac{TCN} based speech separation models for reverberant acoustic conditions. Deformable convolution allows each convolutional layer to have an adaptive \ac{RF}. When used as a replacement for standard convolution in a \ac{TCN} this enables the \ac{TCN} to have a \ac{RF} that can adapt to different reverberant conditions. Using shared weights \cite{furcanext} and dynamic mixing \cite{wavesplit} are also explored as ways to reduce model size and improve performance. A PyTorch library for training deformable 1D convolutional layers as well as a SpeechBrain \cite{speechbrain} \textit{recipe} for reproducing results (cf.~Section \ref{sec:results}) are provided.

The remainder of the paper proceeds as follows. In Section \ref{sec:sigmod} the signal model is discussed. The \ac{DTCN} is introduced in Section \ref{sec:dtcn}. Section \ref{sec:expsetup} discusses the experimental setup, data and baseline systems. Results are given in Section \ref{sec:results}. Section \ref{sec:discussion} provides analysis of the proposed models and conclusions are provided in Section \ref{sec:conclusion}.

\section{Signal Model}\label{sec:sigmod}
A noisy reverberant mixture of $C$ speech signals $s_c[i]$ for discrete sample index $i$ convolved with \acp{RIR} $h_c[i]$ and corrupted by an additive noise signal $\nu[i]$ is defined as
\begin{align}\label{eq:sigmod}
    x[i] &= \sum_{c=1}^{C}h_c[i]*s_c[i]+\nu [i] 
\end{align}
where $\ast$ is the convolution operator. The goal in this work is to estimate the direct speech signal $s_\mathrm{dir,c}[i]$ and remove the reverberant reflections $s_\mathrm{rev,c}[i]$ where
\begin{align}
    x[i] &=\sum_{c=1}^C \left(s_\mathrm{dir,c}[i]+s_\mathrm{rev,c}[i]\right) + \nu[i].
\end{align}

\section{Deformable Temporal Convolutional Separation Network}\label{sec:dtcn}

\subsection{Network Architecture}
\begin{figure}[!ht]
    \centering
    \resizebox{0.95\columnwidth}{!}{
        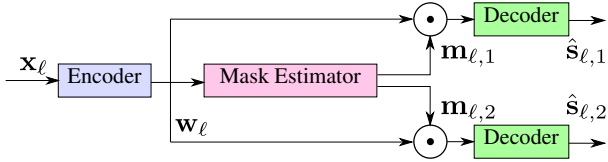
    }
    \caption{Mask-based separation network, exemplary for 2 speakers.}
    \label{fig:masknet}
\end{figure}

The separation network uses a mask-based approach similar to \cite{convtasnet}. The noisy reverberant microphone signal is first segmented into $L_{\vek{x}}$ blocks of length $L_\mathrm{BL}$ with a 50\% overlap defined as
\begin{equation}\label{eq:InputSignalBlock}
    \vek{x}_\ell = \left[x[0.5(\ell-1)L_{\mathrm{BL}}],\ldots, x[0.5(1+\ell) L_{\mathrm{BL}}-1]\right]
\end{equation}
for frame $\ell\in\{1,\ldots,L_\vek{x}\}$. Motivated by \cite{convtasnet,sepformer}, the frames in \eqref{eq:InputSignalBlock} are encoded by a 1D convolutional layer with trainable weights $\mat{B}\in\Real^{L_\mathrm{BL}\times N}$ such that
\begin{equation}\label{eq:encoder}
 \mat{w}_\ell=\mathcal{H}_\mathrm{enc}\left(\vek{x}_\ell\mat{B}\right)
\end{equation}
with a \ac{ReLU} activation function $\mathcal{H}_\mathrm{enc}:\Real^{1\times N} \rightarrow \Real^{ 1\times N}$.
Encoded features $\vek{w}_\ell$ are used as the input to a mask estimation network to produce masks $\vek{m}_{\ell,c}$ for each speaker $c \in \{1,\ldots,C\}$. The masks are then applied to the encoded features using the Hadamard product, i.e.~$\vek{w}_\ell \odot \vek{m}_{\ell,c}$ resulting in $\vek{v}_{\ell,c}$.
The encoded estimate
$\vek{v}_{\ell,c}$ for speaker $c$ can be decoded from the same space back into the time domain using the inverse filter of $\mat{B}$, denoted as $\mat{U}\in\Real^{N\times L_\mathrm{BL}}$, such that
\begin{equation}\label{eq:decoder}
 \hat{\vek{s}}_{\ell,c}= \vek{v}_{\ell,c}\vek{U}
\end{equation}
where $\hat{\vek{s}}_{\ell,c}$ is the estimated clean speech signal for frame $\ell$ in the time domain. These frames are then combined following the overlap-add method.
The entire network model diagram is shown in Fig.~\ref{fig:masknet}. 

\subsection{Mask Estimation Network}
In this subsection, the \ac{DD-Conv} layer is introduced as a replacement for \ac{D-Conv} layers and then the \ac{DTCN} network is described in full.
The mask estimation network consists of \ac{cLN} and a bottleneck \ac{P-Conv} layer which transforms the feature dimension from $N$ to $B$ followed by the \ac{DTCN} which is followed by a \ac{P-Conv} and \ac{ReLU} activation to compute a sequence of masks $\vek{m}_{\ell,:}$ with dimension $C\cdot N$ \cite{convtasnet}.

\subsubsection{Deformable Depthwise Convolution (DD-Conv)}
\begin{figure}[!thb]
    \centering
    \includegraphics{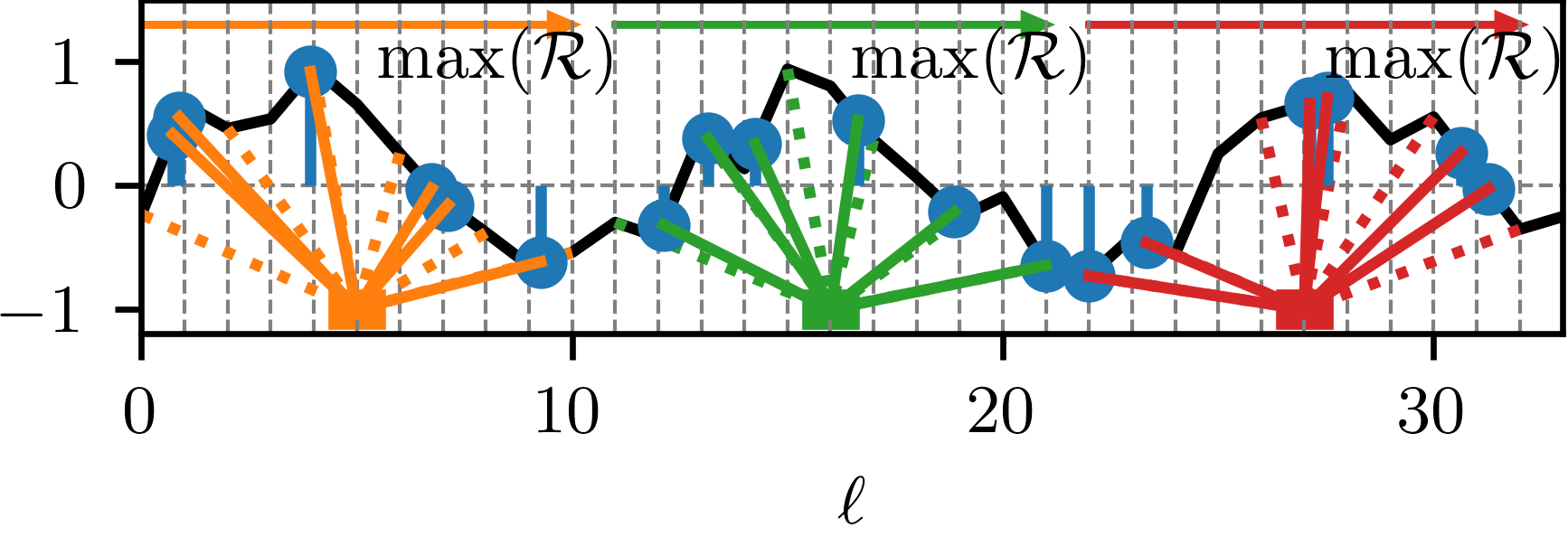}
    \caption{
    Single channel example of 
    Deformable depthwise convolution (bottom) on pseudo-random signal (shown in black) with a kernel size of 6, dilation factor of 2 and stride of 11. $\mathcal{R}$ denotes the \ac{RF} of the kernel. Dotted lines indicate original sampling position of kernel weights before deformation.}
    \label{fig:deform_conv}
\end{figure}

The formulation of \ac{DD-Conv} in this section is adapted from \cite{deformconv} and \cite{1ddconv}. The \ac{D-Conv} operation of kernel size $P$, dilation factor $f$ and convolutional kernel weights for the $g$th channel of an input with $G$ channels denoted $\vek{y}_g\in\Real^{L_\vek{x}}$ at the $\ell$th frame is defined as
\begin{equation} 
\mathcal{D}(\ell,\vek{y}_g,\vek{k}_g)= \sum _{{p}=1}^{P} k_g[{p}]\, y_g[\ell+f\cdot(p-1)].
\end{equation}
The corresponding \ac{DD-Conv} operator with learnable continuous offset of the $p$th kernel weight denoted $\tau_{\ell,p}$ at frame $\ell$ is defined as
\begin{equation} 
\label{eq:DDConv}
\mathcal{C}(\ell,\vek{y}_g,\vek{k}_g ,\tau_{\ell,1:P})= \sum _{{p}=1}^P k_g[{p}]\, y_g[\ell+f\cdot(p-1)+\tau_{\ell,p}]. 
\end{equation}
Note that $\tau_{\ell,p}$ only varies temporally and not across channels. It is feasible to vary these values across channels but in this work to reduce computational complexity offsets are only varied temporally. An illutsration of the \ac{DD-Conv} operation is shown in Figure~\ref{fig:deform_conv}.
To simplify notation let ${\Delta\ell}_p=\ell+f\cdot(p-1)+\tau_{\ell,p}$. Linear interpolation is used to compute values of $y[{\Delta\ell}_p]$ from input sequence $\vek{y}$ such that
\begin{equation} \label{eq:linterpolate}
y[{\Delta\ell}_p]= \sum _{u=\lfloor{\Delta\ell}_p\rfloor}^{\lfloor{\Delta\ell}_p\rfloor+1} 
\max(0,1-|u-{\Delta\ell}_p|)
y[u].
\end{equation}
In practice the interpolation function is designed to constrain the deformable convolutional kernel so it cannot exceed a maximum \ac{RF} of $P\cdot (f-1)+1$ by replacing $u=\lfloor{\Delta\ell}_p\rfloor$ with $u=\min (\lfloor{\Delta\ell}_p\rfloor,\ell+P\cdot(f-1)-1)$ in the bottom of the summation of \eqref{eq:linterpolate}. This constrains the kernel with the benefit of improving interpretability for the overall scope of the \ac{DTCN} described in the following subsection.

\subsubsection{Deformable Temporal Convolutional Mask Estimation Network}
The \ac{DTCN} is formulated in the same way as the Conv-TasNet \ac{TCN} described in \cite{atttasnet}. This implementation deviates slightly from the original Conv-TasNet \cite{convtasnet} by neglecting the \acp{SC} and associated \ac{P-Conv} layers. It was found empirically that these \ac{SC} layers have a negligible impact on performance ($\leq0.1$~dB \ac{SISDR}) while having a significant negative impact on model size ($\approx 35\%$ parameter increase).

\begin{figure}[!t]
    \centering
    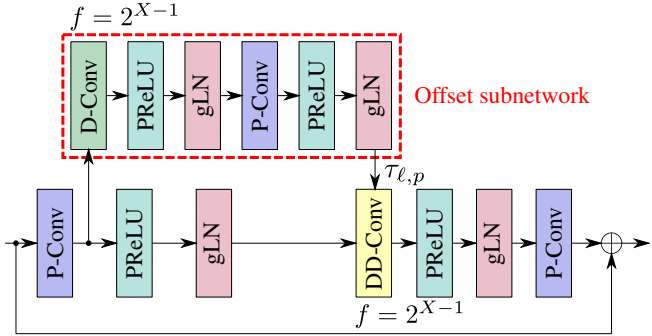
    \caption{Layers inside deformable temporal convolutional blocks. 
    }
    \label{fig:dtcnblocks}
\end{figure}

The \ac{DTCN} is composed of $X\cdot R$ convolutional blocks where $X,R\in\PosInt$ \cite{atttasnet}. Each convolutional block consists of a \ac{P-Conv} which projects the feature dimension from $B$ to $H$, \ac{DD-Conv} that performs a depthwise operation across the $H$ channels and another \ac{P-Conv} layer which projects the feature dimension back to $B$ from $H$. The \ac{DD-Conv} proceeded by \ac{P-Conv} layer forms a \ac{DDS-Conv} structure. \Ac{DS-Conv}, i.e.~a \ac{P-Conv} proceeded by any \ac{D-Conv} layer, is used as a replacement for standard convolutional layers as it is more parameter efficient and mathematically equivalent \cite{convtasnet}. In each convolutional block the \ac{DD-Conv} has an increasing dilation factor $f$ for each additional block in a stack of $X$ blocks as in \cite{atttasnet,convtasnet}. The dilation factor $f$ increases in powers of two through the stack such that $f\in\{1,2,\ldots,2^{X-1}\}$. Note that in \ac{D-Conv} the dilation factor determines the fixed \ac{RF} whereas in the proposed \ac{DD-Conv} the dilation factor defines only the maximum possible \ac{RF} of the kernel. The stack of $X$ convolutional blocks is then repeated $R$ times where the dilation factor is reset to $1$ at the beginning of each stack. Using \ac{SW} for each repeat is experimented with as this significantly reduces the model size similar to \cite{furcanext}. The offsets $\tau_{\ell,p}$ are computed using \ac{DS-Conv} following the initial \ac{P-Conv} in the block, referred to as the offset sub-network. A \ac{PReLU} activation is used at the output as this allows for both negative and positive offsets. Residual connections are applied around each of the convolutional blocks similar to the \ac{TCN} described in \cite{atttasnet}. A schematic of the convolutional blocks is shown in Fig.~\ref{fig:dtcnblocks}.

\section{Experimental Setup}\label{sec:expsetup}
\subsection{Data}
Two datasets are used to evaluate the proposed \ac{DTCN} network. The first is a clean speech mixture corpus known as WSJ0-2Mix \cite{Hershey2016mar}. In WSJ0-2Mix, two speech segments are overlapped at SNRs of $0$ to $5$~dB. The second is a noisy reverberant speech mixture corpus known as WHAMR \cite{WHAMR}. Speech segments in WHAMR are convolved with simulated \acp{RIR} then summed together with mixing SNRs of $0$ to $5$~dB. Ambient noise sources from outdoor pedestrian areas are added at SNRs of $-6$ to $3$~dB (cf. \eqref{eq:sigmod}). \Ac{DM}, i.e.~simulating new training data each epoch, is also experimented with at training time as it has been shown to lead to improved separation performance \cite{sepformer,QDPN}. Speed perturbation training is also performed as part of the \ac{DM} process as described in \cite{sepformer}.

\subsection{Model Configuration}
Feature dimensions $N$, $B$ and $H$, kernel size $P$ and encoder block size $L_\mathrm{BL}$ are fixed:
\begin{equation}
\{N,B,H,P,L_\mathrm{BL}\}=\{512,128,512,3,16\}.
\end{equation} 
These values correspond to the optimal \ac{TCN} network in \cite{convtasnet}. Note $L_\mathrm{BL}$ equates to $2$~ms at 8~kHz.
Five \ac{DTCN} model configurations are evaluated:
\begin{equation}
\{X,R\}\in\{\{3,8\},\{4,6\},\{5,5\},\{6,4\},\{8,3\}\}.
\end{equation}
These configurations are selected as they have a similar or the same number of convolutional blocks to the optimal model configuration in \cite{convtasnet}, i.e.~$\{X,R\}=\{8,3\}$. 
The \ac{SISDR} loss function \cite{tasnet} defined as 
\begin{align} 
\label{eq:DefSISDR}
\mathcal{L}(\hat{\vekt{s}},\vekt{s}_\mathrm{dir})&: 
= \frac{1}{C}\sum_{c=1}^C- 10\log_{10} \frac{\left\Vert \frac{\langle \hat{\vekt{s}}_c,\vekt{s}_{\mathrm{dir},c}\rangle \vekt{s}_{\mathrm{dir},c}}{\Vert \vekt{s}_{\mathrm{dir},c}\Vert^{2}}
\right\Vert^{2}}{\left\Vert\hat{\vekt{s}}_c-\frac{\langle \hat{\vekt{s}}_c,\vekt{s}_{\mathrm{dir},c}\rangle \vekt{s}_{\mathrm{dir},c}}{\Vert \vekt{s}_{\mathrm{dir},c}\Vert^{2}}\right\Vert^{2}}
\end{align}
is used to train the \ac{DTCN} models. \Ac{PIT} is used to solve the speaker permutation problem \cite{upit}.

Two GitHub repositories have been released in conjunction with this work. The first\footnote{URL to dc1d pip repository: \href{https://github.com/jwr1995/dc1d}{\tt github.com/jwr1995/dc1d}} is a Pytorch library for performing 1D deformable convolution. The second\footnote{URL to \ac{DTCN} recipe: \href{https://github.com/jwr1995/DTCN}{\tt github.com/jwr1995/DTCN}} is a model and \textit{recipe} for reproducing our results with the \ac{DTCN} model using the SpeechBrain \cite{speechbrain} framework.

\subsection{Evaluation Metrics}
A number of metrics are used to evaluate the performance of the proposed \ac{DTCN} models. \Ac{SISDR} and \ac{SDR} \cite{LeRoux} are used to measure residual distortion in the signal. \Ac{PESQ} \cite{PESQ} and \ac{ESTOI} \cite{estoi} are used to measure speech quality and intelligibility, respectively. \Ac{SRMR} \cite{srmr} is used to measure residual speech reverberation for the WHAMR corpus.

\section{Results}\label{sec:results}
\begin{figure}[!h]
    \centering
    \adjustbox{trim=0cm 0.26cm 0cm 0.23cm,clip}{
        \includegraphics[width=\columnwidth]{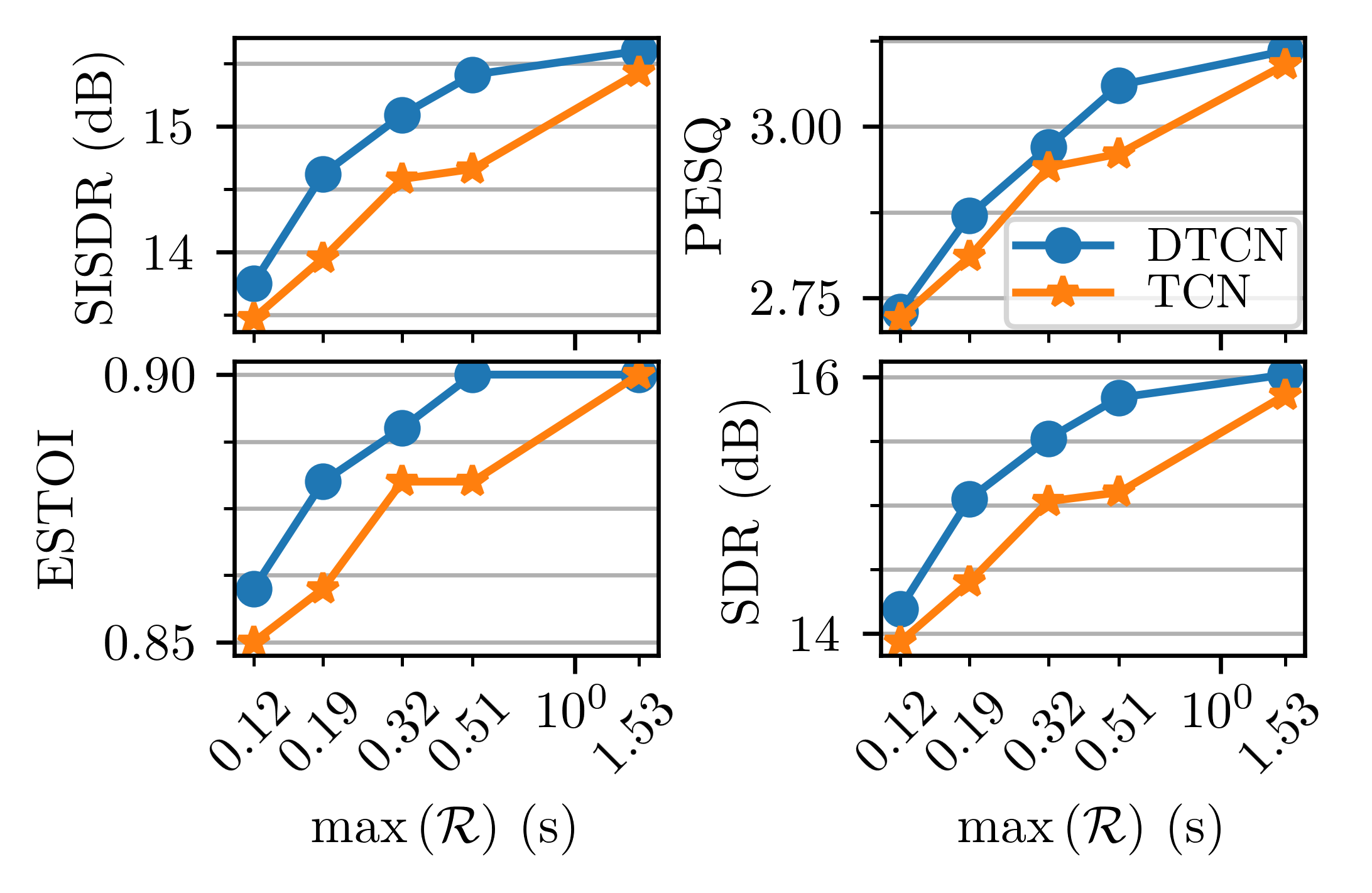}
    }
    \caption{Performance measures over \ac{RF} for WSJ0-2Mix clean speech mixtures.}
    \label{fig:wsj02mix}
\end{figure}

The results for various performance measures against each \ac{DTCN} configuration's \ac{RF} on the clean speech WSJ0-2Mix evaluation are shown in Fig.~\ref{fig:wsj02mix} where they are also compared against their corresponding \ac{TCN} configurations. 
When the model size of the TCN is of the same size by changing $H$ to $532$ the DTCN still outperforms it.
Performance improvements can be seen across all configurations but is more significant with the models which have a \ac{RF} of 0.19s to 0.51s. The improvement in most metrics at the highest \ac{RF} 1.53s is marginal and for the intelligibility metric \ac{ESTOI} the performance is identical to the \ac{TCN}.

\begin{figure}[!h]
    \centering
    \adjustbox{trim=0cm 0.26cm 0cm 0.23cm,clip}{
        \includegraphics[width=\columnwidth]{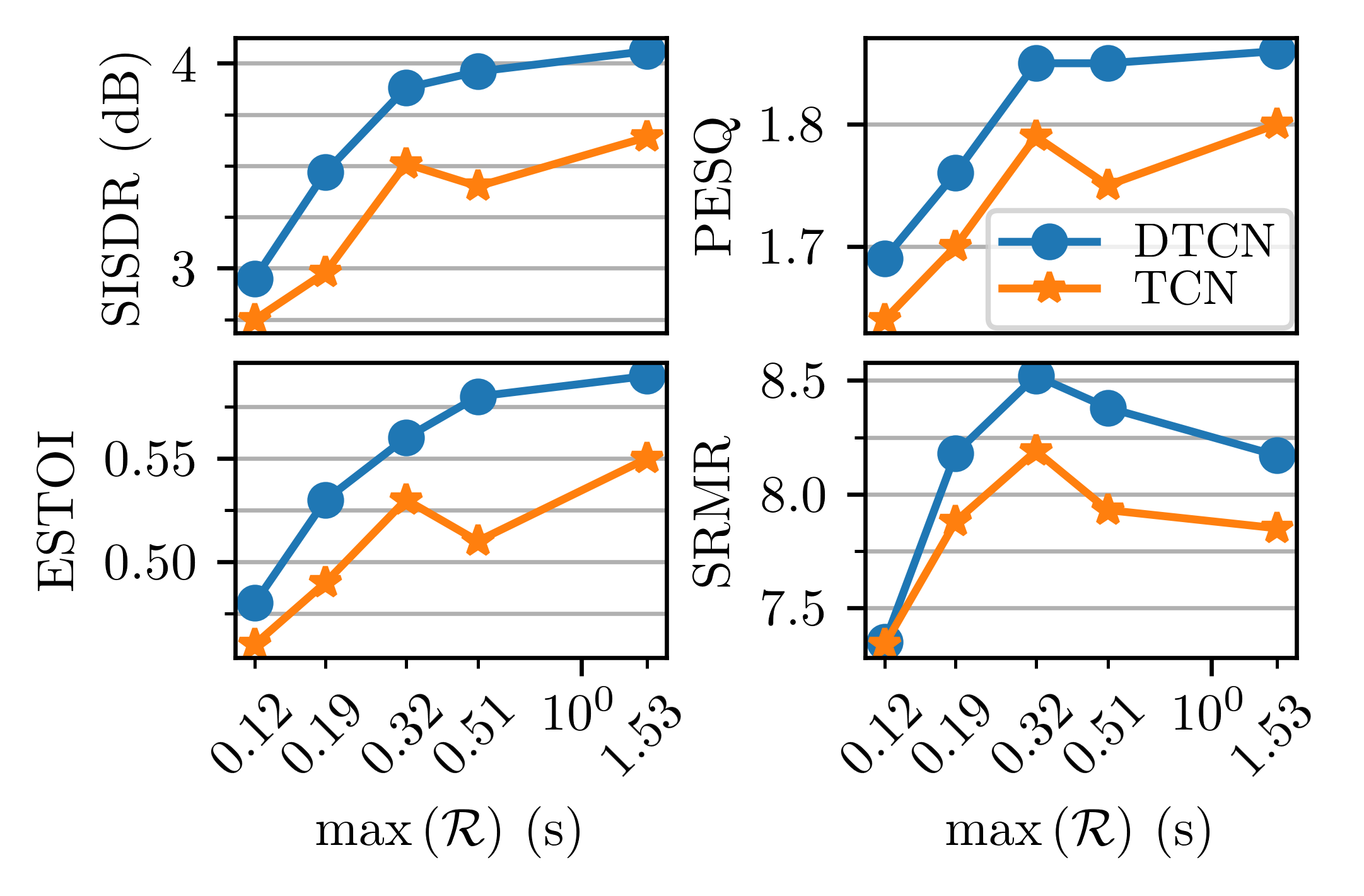}
    }
    \caption{Performance measures over \ac{RF} for WHAMR noisy reverberant speech mixtures.}
    \label{fig:whamr}
\end{figure}
Fig.~\ref{fig:whamr} shows respective results for each \ac{DTCN} configuration's \ac{RF} against the performance measures on noisy reverberant WHAMR data. Note that SDR has been replaced by \ac{SRMR} to provide a measure of reverberation. The \ac{DTCN} again shows improvement over the \ac{TCN} across all measures and model configurations. The performance also increases more consistently as the \ac{RF} increases. The performance convergence seen on the clean speech mixtures in Fig.~\ref{fig:wsj02mix} at the largest \ac{RF}, $\mathcal{R}=1.53$s, is not seen in the results for the noisy reverberant data in Fig.~\ref{fig:whamr}. These findings suggest that deformable convolution is useful in particular for noisy reverberant data.


In Table~\ref{tab:comptab} the proposed \ac{DTCN} model is compared against other speech separation models in terms of size, efficiency and performance.
\begin{table*}[!t]
\caption{Comparison of various \ac{DTCN} models against other speech separation models with varying size and complexity. Dynamic Mixing is abbreviated to DM. Shared weights are abbreviated to SW. Compuational efficiency is expressed in \ac{MACs}. Where possible, all MACs values have been estimated using \textit{thop}\cite{thop} unless a citation is provided.}
\label{tab:comptab}
\centering
\begin{tabular}{c|cc|cc|c|c}
\hline
\textbf{}                           & \multicolumn{2}{c|}{\textbf{WSJ0-2Mix}} & \multicolumn{2}{c|}{\textbf{WHAMR}} &                     &                     \\ \hline
\textbf{Model}                      & $\Delta$\textbf{SISDR}    & $\Delta$\textbf{SDR}    & $\Delta$\textbf{SISDR}  & $\Delta$\textbf{SDR}  & \textbf{Model size} & \textbf{GMACs} \\ \hline
Conv-TasNet \cite{convtasnet} & 15.3                & 15.6              &           9.2 \cite{WHAMR}       &        -        & 5.1M                & 5.2                 \\
Conv-TasNet (w/o SC) & 15.4                & 15.7              & 9.7               & 9.1             & 3.4M                &     3.5               \\
SkiM-KS8 \cite{skim}                & 17.4                & {17.8}              & -                 & -               & 5.9M                & 4.9  \cite{skim}               \\
Tiny-SepformerS-32 \cite{tinysep}   & 15.2               & 16.0             & -                 & -               & 5.3M                & -                   \\
SuDoRM-RF 1.0x++ \cite{sudormrf}    & 17                  & -                 & -                 & -               & 2.7M                &   2.3                 \\
SuDoRM-RF 0.5x \cite{sudormrf}    & 15.4                & -                 & -                 & -               & 1.4M                &        \textbf{1.2 }          \\
FurcaNeXt \cite{furcanext}          & -                   & {18.4}              & -                 & -               & 51.4M               & -                   \\
SepFormer+DM \cite{sepformer}       & {22.3}                &          \textbf{22.4}        & {14.0}                &    \textbf{13.0}             & 26M                 & 69.6 \cite{resepformer}             \\
QDPN+DM \cite{QDPN}                 & \textbf{23.6}                & -                 & \textbf{14.4}              & -               & 200M                &       -             \\ 
\hline
DTCN (proposed)                     &          15.6           &           15.9        & 10.2              & 9.3            & 3.6M                &       3.7            \\
DTCN+DM (proposed)                  &      17.2               &         17.4          &        11.1           &       10.3          & 3.6M                &      3.7              \\
DTCN+SW (proposed)               &       15.0             &       15.3            &      10.0             &       9.3          & \textbf{1.3M}                &         3.7           \\
DTCN+SW+DM (proposed)               &       16.1              &          16.3         &        10.1           &        9.5       & \textbf{1.3M }               &       3.7      \\\hline      
\end{tabular}
\end{table*}
Comparing for model size the proposed \ac{DTCN} outperforms all the Conv-TasNet model baselines including those of equal or larger model size \cite{convtasnet}, and the recurrent SkiM-KS8 model \cite{skim}. When \ac{DM} is used in training, the \ac{DTCN} outperforms the much larger convolutional SuDo-RM-RF 1.0x++ model \cite{sudormrf}. Using \ac{SW} reduces the model size by two-thirds but is still able to give comparable performance to the SuDoRM-RF 0.5x model of similar size and much-improved performance when \ac{DM} is also used.

\section{Analysis}\label{sec:discussion}
In the following the offset values of the best-performing model configuration $\{X,R\}={8,3}$ are analysed with the aim to provide insight as to how temporal offsets $\tau_{\ell,p}$ in \eqref{eq:DDConv}, cf.~also Fig.~\ref{fig:deform_conv}, behave relative to one another. 
The $2$nd convolutional block of the $2$nd repeat in the \ac{DTCN} (i.e.~the $10$th block overall) was selected for analysis as it was found to have the highest average offset variance over the WHAMR evaluation set. The motivation for this choice is that it is assumed that blocks with offsets of larger variances are more indicative of the benefits of using deformable convolution.
\begin{figure}[!ht]
    \centering
    \includegraphics{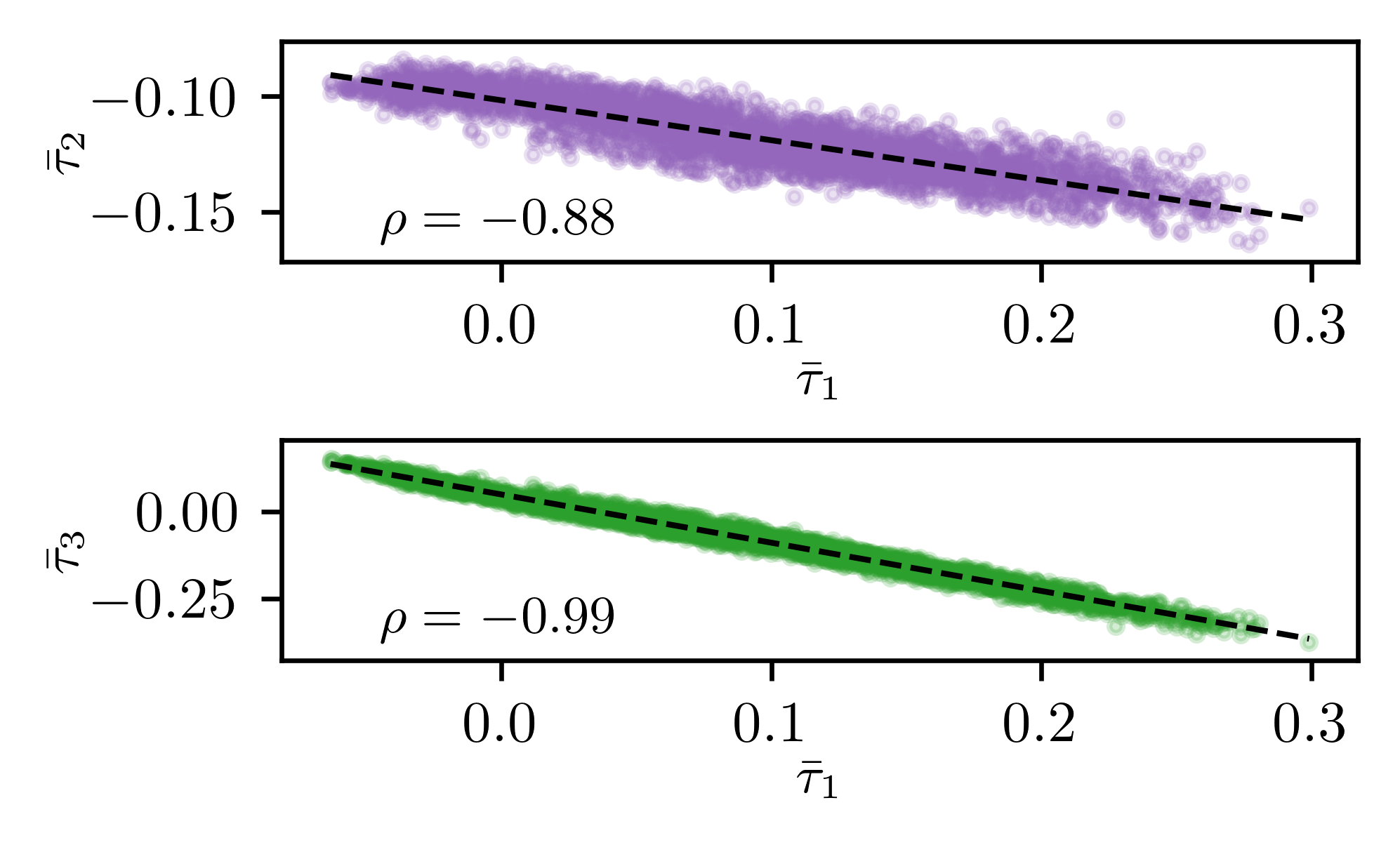}
    \caption{Mean offset values $\bar{\tau}_2$ (top) and $\bar{\tau}_3$ (bottom) of the $10$th convolutional block of the \ac{DTCN} model plotted against the mean offset value of the first kernel weight $\bar{\tau_1}$ for each example in the WHAMR evaluation set. Pearson correlation coefficients are denoted with $\rho$. Dashed black line indicates line of best fit.}
    \label{fig:tauvstau}
\end{figure}
A correlation analysis was performed between each of three offset values averaged per utterance $\bar{\tau}_p$, corresponding to the three kernel weight positions. Fig.~\ref{fig:tauvstau} shows scatter plots for the mean of the middle and outermost offsets denoted $\bar{\tau}_2$ and $\bar{\tau}_3$, respectively, against the mean offset value of the first kernel sample point $\bar{\tau}_1$ for every example in the evaluation set. A strong negative correlation ($\rho=-0.99$) can be observed between $\bar{\tau}_1$ and $\bar{\tau}_3$ indicating that the deformation is causing the \ac{RF} of the kernel to shrink and grow more than shifting its focal point. A less strong negative correlation ($\rho=-0.88$) was found between $\bar{\tau}_1$ and $\bar{\tau}_2$ indicating similar behaviour. The comparison of $\bar{\tau}_2$ against $\bar{\tau}_3$ is omitted from Fig~\ref{fig:tauvstau} for brevity but these mean offset values were found to have a positive correlation of $\rho=0.81$.

\section{Conclusion}\label{sec:conclusion}
In this paper deformable convolution was proposed as a method to improve \acp{TCN} for noisy reverberant speech separation. It was shown that the \ac{DTCN} model is particularly useful for noisy reverberant conditions as performance increases were less consistent in the case of anechoic speech separation with a sufficiently large receptive field. 
Using shared weights and dynamic mixing led to further performance improvements resulting in a small model size for the \ac{DTCN} compared to other separation models which give comparable performance. 
Finally, it was shown that the \ac{DTCN} offsets vary the size of the receptive field of convolutional blocks in the network relative to the input data.

\bibliographystyle{IEEEbib} 
\bibliography{refs}

\end{document}